\documentclass[12pt,a4paper]{article}

\usepackage{amsmath,amsthm,amssymb,amscd,a4wide}

\usepackage{dsfont}
\usepackage{mathrsfs}

\newcommand{\ud}{\mathrm{d}}
\newcommand{\ui}{\mathrm{i}}
\newcommand{\ue}{\mathrm{e}}

\newcommand{\SU}{\mathrm{SU}}
\newcommand{\SO}{\mathrm{SO}}

\newcommand{\Stwo}{\mathrm{S}^2}

\newcommand{\vn}{\boldsymbol{n}}
\newcommand{\vp}{\boldsymbol{p}}

\newcommand{\vx}{\boldsymbol{x}}
\newcommand{\vy}{\boldsymbol{y}}
\newcommand{\vA}{\boldsymbol{A}}
\newcommand{\vB}{\boldsymbol{B}}
\newcommand{\vC}{\boldsymbol{C}}
\newcommand{\vE}{\boldsymbol{E}}
\newcommand{\vF}{\boldsymbol{F}}

\newcommand{\valpha}{\boldsymbol{\alpha}}
\newcommand{\vsigma}{\boldsymbol{\sigma}}
\newcommand{\vxi}{\boldsymbol{\xi}}

\newcommand{\cH}{{\hat H}}

\newcommand{\rz}{{\mathbb R}}
\newcommand{\nz}{{\mathbb N}}
\newcommand{\kz}{{\mathbb C}}

\DeclareMathOperator{\mtr}{tr}
\DeclareMathOperator{\vol}{vol}

\DeclareMathOperator{\grad}{grad}
\DeclareMathOperator{\rot}{rot}

\newcommand{\dpr}{\partial}

\newcommand{\eins}{\mathds{1}}

\numberwithin{equation}{section}

\newtheorem{theorem}{Theorem}[section]

\theoremstyle{definition}

\newcommand{\Osc}{\cal{O}_{\mathrm{sc}}}
\newcommand{\Oinv}{\cal{O}_{\mathrm{inv}}}

\begin{document}

\thispagestyle{empty}

\noindent
ULM-TP/04-2\\
February 2004\\

\vspace*{1cm}

\begin{center}

{\LARGE\bf Zitterbewegung and semiclassical observables \\[5mm]
for the Dirac equation} \\
\vspace*{3cm}
{\large Jens Bolte}%
\footnote{E-mail address: {\tt jens.bolte@physik.uni-ulm.de}}
{\large and Rainer Glaser}%
\footnote{E-mail address: {\tt rainer.glaser@physik.uni-ulm.de}}

\vspace*{1cm}

Abteilung Theoretische Physik\\
Universit\"at Ulm, Albert-Einstein-Allee 11\\
D-89069 Ulm, Germany 
\end{center}

\vfill

\begin{abstract}
In a semiclassical context we investigate the Zitterbewegung of relativistic 
particles with spin 1/2 moving in external fields. It is shown that the 
analogue of Zitterbewegung for general observables can be removed to 
arbitrary order in $\hbar$ by projecting to dynamically almost invariant 
subspaces of the quantum mechanical Hilbert space which are associated with 
particles and anti-particles. This not only allows to identify observables 
with a semiclassical meaning, but also to recover combined classical
dynamics for the translational and spin degrees of freedom. Finally, we
discuss properties of eigenspinors of a Dirac-Hamiltonian when these are
projected to the almost invariant subspaces, including the phenomenon of 
quantum ergodicity.
\end{abstract}

\newpage

\section{Introduction}
\label{sec:intro}
When Dirac had introduced his relativistic quantum theory for spin-1/2
particles it became soon clear that despite its overwhelming success in
producing the correct hydrogen spectrum, including fine structure, this 
theory was plagued with a number of apparent inconsistencies as, e.g., 
Klein's paradox \cite{Kle29}. Schr\"odinger observed \cite{Sch30} that 
the (free) time evolution of the naive position operator, that he had taken 
over from non-relativistic quantum mechanics, contained a contribution 
without any classical interpretation. Since this term was rapidly oscillating 
he introduced the notion of {\it Zitterbewegung (trembling motion)}. 
Schr\"odinger wanted to explain this effect as being caused by spin. Later, 
however, it became clear that all paradoxa stemmed from the coexistence of 
particles and anti-particles in Dirac's theory, and that only a quantum field 
theoretic description, predicting pair creation and annihilation, could cure 
these problems in a fully satisfactory manner.

It nevertheless turned out that relativistic quantum mechanics was capable 
of describing many important physical phenomena; e.g., in atomic physics it 
serves to explain spectra of large atoms and in nuclear physics it allows 
to construct models of nuclei. Within the context of this theory one can 
remove the Zitterbewegung of free particles by introducing modified 
position operators that are associated with either only particles or only 
anti-particles. These operators are free from the particle/anti-particle 
interferences that cause the Zitterbewegung; for details see e.g.\ 
\cite{Tha92}. In the case of interactions (with external potentials) the 
Zitterbewegung cannot be exactly eliminated. It is, however, possible to 
devise an asymptotic construction, e.g. based on the Foldy-Wouthuysen 
transformation. This involves a non-relativistic expansion, and when applied 
to the Dirac equation reproduces in leading order the Pauli equation. The 
genuinely relativistic coexistence of particles and anti-particles is 
therefore removed in a natural way.

Similar constructions are possible in a semiclassical context while
maintaining the relativistic level of description. The decoupling of particles
from anti-particles then proceeds asymptotically order by order in powers
of the semiclassical parameter $\hbar$. In this direction several approaches 
have been developed recently, aiming at semiclassical expansions for
scattering phases \cite{BruNou99,BruRob99} or at recovering the Thomas
precession of the spin \cite{Spo00}, see also \cite{Cor01,Teu03}. In other
semiclassical approaches trace formulae of the Gutzwiller type have been
derived \cite{BolKep98,BolKep99a}, see also \cite{Kep03}. Here 
our goal is to introduce a general semiclassical framework for the Dirac 
equation, and in particular to identify a sufficiently large algebra of 
quantum mechanical observables that can be assigned a classical analogue 
in a way that is consistent with the dynamics. And in contrast to previous
studies we do not construct semiclassical (unitary) Foldy-Wouthuysen
operators but use the associated projection operators, thus avoiding
ambiguities in the choice of the unitaries.

The outline of this paper is as follows. In section~\ref{freeDirac} we
recall the phenomenon of Zitterbewegung of free relativistic particles,
emphasising that this effect can be removed through an introduction of
suitably modified position operators. The following section is devoted
to the construction of projection operators that semiclassically separate
particles from anti-particles in the presence of an interaction with external
potentials. Quantum observables with a semiclassical meaning are introduced 
in section~\ref{sec:SclObserv}, and a characterisation of such observables 
that maintain this property under their time evolution is provided. 
Section~\ref{sec:TimeEvol} contains a discussion of the classical dynamics 
emerging from the quantum dynamics generated by a Dirac-Hamiltonian in the 
semiclassical limit. It is shown that a combined time evolution of the 
translational and the spin degrees of freedom arises in the form of a 
so-called skew-product. Finally, in section~\ref{sec:SCeigen} we discuss
to what extent eigenspinors of a Dirac-Hamiltonian can be associated
with either particles or anti-particles. We also show that an ergodic
behaviour of the combined classical dynamics implies quantum ergodicity, 
i.e. a semiclassical equidistribution on particle and anti-particle energy 
shells of eigenspinor-projections to the respective energy shells.
\section{Zitterbewegung for the free motion}
\label{freeDirac}

Since its early discovery by Schr\"odinger Zitterbewegung has mostly been
discussed in the context of the Dirac equation for a free relativistic
particle of mass $m$ and spin 1/2, whose Hamiltonian reads
\begin{equation} 
\label{eq:freeDiracOp}
 \cH_{0}:= -\ui\hbar c \, \valpha\cdot\nabla + \beta m c^2 \ .
\end{equation}
The relations $\alpha_k\alpha_l +\alpha_l\alpha_k =2\delta_{kl}$ and 
$\alpha_k\beta+\beta\alpha_k =0$ ($k,l=1,2,3$) defining the Dirac-algebra 
are realised by the $4\times 4$ matrices
\begin{equation*}
 \valpha = \left( \begin{array}{cc} 0 & \vsigma \\ \vsigma & 0 
 \end{array}\right) 
 \qquad \text{and} \qquad 
 \beta = \left( \begin{array}{cc} \eins_2 & 0 \\ 0& - \eins_2 \end{array} 
 \right) \ ,
\end{equation*}
where $\vsigma=(\sigma_1,\sigma_2,\sigma_3)$ is the vector of Pauli spin 
matrices and $\eins_n$ denotes the $n\times n$ unit matrix. The operator 
(\ref{eq:freeDiracOp}) is essentially self-adjoint on the dense domain 
$C_0^\infty(\rz^3)\otimes\kz^4$ in the Hilbert space 
$\mathscr{H}:=L^2(\rz^3) \otimes \kz^4$; its spectrum is absolutely 
continuous and comprises of $(-\infty,-mc^2)\cup (mc^2,\infty)$, see e.g.
\cite{Tha92}. The time evolution (free motion)
\begin{equation*}
 \hat U_0(t) := \ue^{-\frac{\ui}{\hbar}\hat H_0 t}
\end{equation*}
hence is unitary.

In his seminal paper \cite{Sch30} Schr\"odinger solved the free time
evolution of the standard position operator $\hat{\vx}$, whose components
are defined as multiplication operators on a suitable domain in 
$L^2(\rz^3)\otimes\kz^4$. His intention was to resolve the apparent paradox 
that the spectrum of the velocity operator resulting from the Heisenberg 
equation of motion for the standard position operator, 
$\dot{\hat{\vx}}(t)=c\valpha (t)$, consists of $\pm c$, whereas the classical
velocity of a free relativistic particle reads 
$c^2\vp/\sqrt{\vp^2 c^2 +m^2 c^4}$ and thus is smaller than $c$ in magnitude. 
Integrating the equations of motion Schr\"odinger found
\begin{equation} 
\label{eq:XofT}
\begin{split}
 \hat{\vx}(t) &= \hat U_0(t)^\ast\,\hat{\vx}\,\hat U_0(t) \\
              &= \hat{\vx}(0) + c^2 \hat{\vp} \cH_0^{-1} t + 
                 \frac{\hbar}{2 \ui} \cH_0^{-1} \bigl( 
                 \ue^{\frac{2 \ui}{\hbar} \cH_0 t} -1 \bigr) \hat{\vF} \ .
\end{split}
\end{equation}
The first two terms of this result exactly correspond to the respective 
classical dynamics of a free relativistic particle. The third term, however, 
contains the operator
\begin{equation*} 
 \hat{\vF} := c\valpha - c^2 \hat{\vp} \cH_0^{-1} \ ,
\end{equation*}
which is well defined since zero is not in the spectrum of $\cH_0$ and thus 
$\cH_0^{-1}$ is a bounded operator. The quantity $\hat{\vF}$ expresses the 
difference between the velocity operator and the quantisation of the classical
velocity. This additional term introduces a rapidly oscillating time 
dependence and hence was named {\it Zitterbewegung (trembling motion)} by 
Schr\"odinger.

After Schr\"odinger's work \cite{Sch30} the origin of the Zitterbewegung
was traced back to the coexistence of particles and anti-particles in
relativistic quantum mechanics. Therefore, this effect can be removed by 
projecting the standard position operator to the particle and anti-particle 
subspaces of $\mathscr{H}$, respectively. To this end one introduces the 
projection operators
\begin{equation} 
\label{eq:FreeSpecProj}
 \hat P_0^\pm:= \frac{1}{2} \left( \eins_{\mathscr{H}}\pm |\cH_0|^{-1}\cH_0 
 \right) \ , 
\end{equation} 
with $| \cH_0 | := (\cH_0^\ast \cH_0)^{1/2}$, fulfilling 
$\hat P_0^+\hat P_0^-=0$ and $\hat P_0^+ +\hat P_0^-=\eins_{\mathscr{H}}$. 
The time evolution of the projected position operators 
\begin{equation} 
\label{eq:ProjPosOp}
 \hat{\vx}^\pm_0 := \hat P_0^\pm \, \hat{\vx} \, \hat P_0^\pm
\end{equation} 
is then given by (see \cite{Tha92})
\begin{equation*} 
 \hat{\vx}^\pm_0 (t) = 
 \hat{\vx}^\pm_0 (0) + c^2 \hat{\vp} \cH_0^{-1} t \, \hat P_0^\pm \ .
\end{equation*}
It hence exactly corresponds to the respective classical expression;
and this is true for both particles and anti-particles. The interpretation 
of this observation is obvious: Particles and anti-particles show noticeably 
different time evolutions and the Zitterbewegung in (\ref{eq:XofT}) represents
the interference term \cite{Tha92}
\begin{equation*} 
 \hat P_0^+\hat{\vx}\hat P_0^- + \hat P_0^-\hat{\vx}\hat P_0^+ =
 \frac{\ui\hbar}{2}\,\hat H_0^{-1}\,\hat{\vF}
\end{equation*}
that is absent in a classical description.

We remark that the projected position operators (\ref{eq:ProjPosOp}) differ
from the Newton-Wigner position operators introduced in \cite{NewWig49}. 
Whereas the latter also respect the splitting of the Hilbert space into 
particle and anti-particle subspaces, they arise from the standard position 
operator by unitary transformations. The Newton-Wigner operators are unique 
in the sense that they possess certain natural localisation properties 
\cite{Wig62}. However, it is well known that in relativistic quantum mechanics
no position operators exist that leave the particle and anti-particle 
subspaces invariant, share the natural localisation properties and do not 
violate Einstein causality, see e.g. the discussion in \cite{Tha92}. Thus 
even for free particles no complete quantum-classical correspondence exists. 
The goal we want to achieve in the following hence is to promote a receipt that 
allows to separate particles from anti-particles in a semiclassical fashion: 
The Hilbert space is split into mutually orthogonal subspaces and observables
are projected to these subspaces. Within these subspaces one can then 
employ semiclassical techniques and set up quantum-classical correspondences. 
In the next section we are going to extend this concept to particles 
interacting with external fields. In this situation even a separation of 
$\mathscr{H}$ into subspaces that are associated with particles and 
anti-particles, respectively, can only be achieved asymptotically (in the 
semiclassical limit).

\section{Semiclassical projection operators}
\label{sec:SclCalc}

We now consider the Dirac equation 
\begin{equation*}
 \ui\hbar\frac{\partial}{\partial t}\psi(t,\vx) = \hat H \psi(t,\vx)
\end{equation*}
for a particle of mass $m$ and charge $e$ coupled to external time-independent
electromagnetic fields $\vE(\vx)=-\grad\phi(\vx)$ and $\vB(\vx)=\rot\vA(\vx)$.
The Hamiltonian therefore reads
\begin{equation}
\label{eq:DiracHam}
 \hat H = c\valpha \cdot \Bigl( \frac{\hbar}{\ui}\nabla -
 \frac{e}{c}\vA(\vx) \Bigr) +\beta mc^2 + e\phi(\vx) \ .
\end{equation}
For convenience we restrict our attention to smooth potentials, in which case
the operator (\ref{eq:DiracHam}) is known to be essentially self-adjoint
on the domain $C_0^\infty(\rz^3)\otimes\kz^4$, see e.g. \cite{Tha92},
and hence defines a unitary time evolution
\begin{equation*}
 \hat U(t) := \ue^{-\frac{\ui}{\hbar}\hat H t} \ .
\end{equation*}

In order to separate particles and anti-particles also in the presence of
electromagnetic fields one would like to construct projection operators 
analogous to (\ref{eq:FreeSpecProj}) that commute with the Hamiltonian.
This, however, is not possible as can be seen by considering operators
in a phase-space representation. Here we choose the Wigner-Weyl calculus
in which an operator $\hat B$ on $\mathscr{H}$ is defined in terms of
a function $B(\vx,\vp)$ on phase space taking values in the $4\times 4$
matrices,
\begin{equation} 
\label{eq:WeylQuant}
 (\hat B \psi)(\vx) = \frac{1}{(2\pi\hbar)^3} \iint_{\rz^3\times\rz^3} 
 \ue^{\frac{\ui}{\hbar}\vp\cdot(\vx-\vy)} \, 
 B \Bigl( \frac{\vx+\vy}{2},\vp\Bigr) \, \psi(\vy) \ \ud y \, \ud p  \ .
\end{equation}
This operator is well defined on Dirac-spinors 
$\psi\in\mathscr{S}(\rz^3)\otimes\kz^4$, if the matrix valued Weyl symbol 
$B(\vx,\vp)$ is smooth. For this and further details of the Weyl calculus 
see \cite{Fol89,Rob87,DimSjo99}. Introducing then, e.g., the symbol
\begin{equation}
\label{eq:DiracSymb}
 H (\vx,\vp) = c\valpha \cdot \Bigl( \vp - \frac{e}{c}\vA(\vx) \Bigr) +
 \beta mc^2 + e\phi(\vx) \ ,
\end{equation}
one can represent the Hamiltonian (\ref{eq:DiracHam}) as a Weyl operator 
(\ref{eq:WeylQuant}). For each point $(\vx,\vp)$ in phase space the symbol
(\ref{eq:DiracSymb}) is a hermitian $4\times 4$ matrix with the two
doubly degenerate eigenvalues
\begin{equation}
\label{eq:DiracSymbEv}
 h_\pm (\vx,\vp) = e\phi(\vx) \pm \sqrt{(c\vp-e\vA(\vx))^2 +m^2c^4} \ .
\end{equation}
These functions can immediately be identified as the classical Hamiltonians
of particles and anti-particles, respectively, without spin. Associated with
these eigenvalues are the projection matrices 
\begin{equation} 
\label{eq:EigenProj}
 \Pi^\pm_0 (\vx,\vp) = \frac{1}{2} \left( \eins_4 \pm 
 \frac{\valpha\cdot (c\vp - e\vA(\vx)) + \beta mc^2}
 {\sqrt{(c\vp - e\vA(\vx))^2 + m^2 c^4}} \right)  
\end{equation}
onto the respective eigenspaces in $\kz^4$. In the absence of potentials 
the matrix valued functions (\ref{eq:EigenProj}) are independent of $\vx$ 
and their Weyl quantisations according to (\ref{eq:WeylQuant}) yield the 
projectors (\ref{eq:FreeSpecProj}). One could hence be tempted to view
the Weyl quantisations $\hat\Pi_0^\pm$ of the symbols (\ref{eq:EigenProj}) 
as appropriate substitutes for the operators (\ref{eq:FreeSpecProj}) 
in the general case. Indeed, if the partial derivatives of $\vA(\vx)$ are 
bounded, the quantities (\ref{eq:EigenProj}) and all their partial derivatives
are bounded smooth functions of $(\vx,\vp)$ such that the operators 
$\hat\Pi_0^\pm$ are self-adjoint and bounded on the Hilbert space 
$\mathscr{H}$ \cite{CalVai71}. The fact that the symbols (\ref{eq:EigenProj}) 
depend on $\vx$, however, implies that $\hat\Pi_0^\pm$ are not projection 
operators, but satisfy \cite{EmmWei96}
\begin{equation} 
\label{eq:NoProj}
 \bigl( \hat\Pi_0^\pm \bigr)^2 - \hat\Pi_0^\pm = O(\hbar) \ .
\end{equation}
Moreover, $\hat\Pi_0^\pm$ do not commute with the Hamiltonian.

The error on the right-hand side of (\ref{eq:NoProj}) can be improved by 
adding a suitable term of order $\hbar$ to the symbol (\ref{eq:EigenProj}).
To guarantee that the Weyl quantisation of this symbol leads to a bounded 
operator, we now demand that the potential $\phi(\vx)$ and all its partial 
derivatives are bounded, and that $\vA(\vx)$ grows at most like some power 
$|\vx|^K$. In this case the quantity
\begin{equation} 
\label{eq:OrderFct}
 \varepsilon(\vx,\vp) := \sqrt{(c\vp-e\vA(\vx))^2 +m^2c^4}
\end{equation}
may serve as a so-called order function for the symbol (\ref{eq:DiracSymb})
and the construction of \cite{BolGla02} applies. After quantisation of the
corrected symbol this leads to an operator that is a projector up to an 
error of order $\hbar^2$. This procedure can be repeated to an arbitrary 
order $\hbar^N$, see \cite{EmmWei96,BruNou99}. One hence obtains almost 
projection operators $\hat\Pi^\pm$ that almost commute with the Hamiltonian,
\begin{equation} 
\label{eq:SemiclCommute}
 \bigl( \hat\Pi^\pm \bigr)^2 - \hat\Pi^\pm = O(\hbar^\infty) \qquad \text{and}
 \qquad [\hat H,\hat\Pi^\pm] = O(\hbar^\infty) \ ,
\end{equation}
meaning that the operator norm of the above expressions is smaller than 
any power of $\hbar$. Furthermore,
\begin{equation} 
\label{eq:SemiclResUn}
 \hat\Pi^+\hat\Pi^- = O(\hbar^\infty) \qquad\text{and}\qquad
 \hat\Pi^+ + \hat\Pi^- = \eins_{\mathscr{H}} + O(\hbar^\infty) \ .
\end{equation}
As a consequence of $\hat\Pi^\pm$ being almost projectors, their spectrum
is concentrated around zero and one. The standard Riesz projection formula
therefore allows to construct the genuine orthogonal projectors
\begin{equation*}
 \hat P^\pm := \frac{1}{2\pi\ui}\int_{|\lambda-1|=\frac{1}{2}} \bigl( 
 \hat\Pi^\pm - \lambda \bigr)^{-1} \ \ud\lambda \ ,
\end{equation*}
which also almost commute with $\hat H$ and fulfill (\ref{eq:SemiclResUn}).

Since the above construction is based on the separation into particles and 
anti-particles on a classical level (\ref{eq:DiracSymbEv}) and is then 
extended order by order in $\hbar$, one can view the subspaces 
$\mathscr{H}^\pm:=\hat P^\pm\mathscr{H}$ as being semiclassically associated 
with particles and anti-particles, respectively. Moreover, due to the 
relation (\ref{eq:SemiclCommute}) these subspaces are almost invariant with 
respect to the time evolution generated by $\hat H$, i.e.
\begin{equation*} 
 \hat U(t)\,\hat P^\pm\,\psi - \hat P^\pm\,\hat U(t)\,\psi 
 = O(t\hbar^\infty) \qquad\text{for all}\ \psi\in\mathscr{H}\ .
\end{equation*}
Hence, up to a small error every spinor in one of the subspaces 
$\mathscr{H}^\pm$ remains under the time evolution within this subspace; 
and this is true for semiclassically long times $t\ll\hbar^{-N}$, where $N$ 
can be chosen arbitrarily large. This result is in agreement with the 
(heuristic) physical picture that particles interact with anti-particles
via tunneling; the latter is a genuine quantum process with pair production 
and annihilation rates that are exponentially small in $\hbar$. Related to 
this observation is the fact that eigenspinors of the Hamiltonian can (only) 
almost be associated with particles or anti-particles: If 
$\psi_n\in\mathscr{H}$ is an eigenspinor, $\hat H\psi_n =E_n\psi_n$, its 
projections $\hat P^\pm\psi_n$ are in general only almost eigenspinors 
(quasimodes), i.e.
\begin{equation*} 
 ( \hat H - E_n ) \hat P^\pm\psi_n = O(\hbar^\infty) \ .
\end{equation*}
Thus the discrete spectrum of the Hamiltonian cannot truly be divided
into a particle and an anti-particle part. Further consequences will be
investigated in section \ref{sec:SCeigen}.

The above discussion raises the question whether the semiclassical projectors 
$\hat P^\pm$ can (at least semiclassically) be related to spectral projections
of the Hamiltonian. According to (\ref{eq:FreeSpecProj}), in the case without 
potentials the operators $\hat P_0^\pm$ are obviously the spectral projectors 
to $(-\infty,-mc^2)$ and $(mc^2,\infty)$, which are exactly the (absolutely 
continuous) spectral stretches associated with particles and anti-particles, 
respectively. In the presence of potentials, depending on their behaviour at 
infinity, there exist constants $E_+> E_-$ such that the spectrum inside 
$(E_-,E_+)$ is discrete and absolutely continuous outside; e.g., if the 
potentials and their derivatives vanish at infinity one finds 
$E_\pm = \pm mc^2$, see \cite{Tha92}. For $E\in (E_-,E_+)$ not in the spectrum
of $\hat H$ we then denote the spectral projectors to $(-\infty,E)$ and 
$(E,\infty)$ by $\hat P^-_E$ and $\hat P^+_E$, respectively. For the 
semiclassical considerations to follow we also assume that 
\begin{equation*}
 \sup h_-(\vx,\vp) < E < \inf h_+ (\vx,\vp) \ ;
\end{equation*}
in particular, it is required here that the classical particle and 
anti-particle Hamiltonians are separated by a gap as $(\vx,\vp)$ runs over 
phase space. Therefore, the two classical energy shells
\begin{equation*}
 \Omega_E^\pm = \{ (\vx,\vp);\ h_\pm (\vx,\vp) = E \}
\end{equation*}
must not intersect nor touch. On the submanifold $\vp = \frac{e}{c}\vA(\vx)$ 
this condition requires that the variation of the potential $\phi(\vx)$ is 
strictly restricted by $2mc^2$. In such a situation the semiclassical 
projectors $\hat P^\pm$ are semiclassically close to the spectral projectors,
\begin{equation*}
 \| \hat P^\pm - \hat P^\pm_E \| = O (\hbar^\infty) \ . 
\end{equation*}
A full proof of this statement can be found in \cite{BolGla02}.

\section{Semiclassical observables}
\label{sec:SclObserv}

The example of the standard position operator for the free particle and
its Zitterbewegung demonstrates that not all quantum observables possess
a direct (semi-) classical interpretation. Only the diagonal blocks
with respect to the projections $\hat P_0^\pm$ allow for establishing
quantum-classical correspondences. This observation applies in particular 
to dynamical questions because only after projection there exists an
unambiguous classical Hamiltonian (\ref{eq:DiracSymbEv}). Following the 
discussion in the previous section, these remarks carry over to the general 
case appropriately. Our aim therefore is to identify a sufficiently large 
class of observables with a semiclassical meaning. We moreover require this 
class to be invariant with respect to the quantum time evolution, i.e., 
$\hat B$ being of this class should imply that 
$\hat B(t)=\hat U(t)^\ast\hat B\hat U(t)$ is of the same type. 

A natural starting point in the present context is to choose the Weyl
representation (\ref{eq:WeylQuant}) for operators on $\mathscr{H}$. It is
well known that this is possible for a fairly large class of operators, 
namely those that are continuous from 
$\mathscr{S}(\rz^3\times\rz^3)\otimes\kz^4$ to 
$\mathscr{S}'(\rz^3\times\rz^3)\otimes\kz^4$, see e.g. \cite{Fol89,DimSjo99}. 
In order to achieve manageable algebraic properties, however, one must 
restrict the class of operators further. For convenience we here choose Weyl 
quantisations of smooth symbols $B(\vx,\vp)$ that are, along with all of 
their partial derivatives, bounded on $\rz^3\times\rz^3$. This restriction 
leads to bounded operators on $\mathscr{H}$, see \cite{CalVai71}. We admit 
$\hbar$-dependent symbols, since even when restricting to quantisations 
$\hat B$ of $\hbar$-independent symbols $B(\vx,\vp)$ their time evolutions 
$\hat B(t)$ will necessarily acquire an $\hbar$-dependence in their symbols 
$B(t)(\vx,\vp;\hbar)$. But for the purpose of convenient semiclassical 
asymptotics we restrict the $\hbar$-dependence in a particular way,
\begin{equation} 
\label{eq:ClassSymb}
 B(\vx,\vp;\hbar) \sim \sum_{k=0}^\infty \hbar^k \, B_k(\vx,\vp) \ .
\end{equation} 
The asymptotic expansion is to be understood in the sense that
\begin{equation}
\label{eq:AsympExpand}
 \Bigl\| \dpr_x^\alpha \dpr_p^\beta \Bigl( B(\vx,\vp;\hbar) - 
 \sum_{k=0}^{N-1} \hbar^k \, B_k (\vx,\vp) \Bigr) \Bigr\|_{4 \times 4} 
 \le C_{\alpha,\beta}^{(N)}\, \hbar^N 
\end{equation} 
holds for all $(\vx,\vp)\in\rz^3\times\rz^3$ and all $N\in\nz$ uniformly 
in $\hbar$ as $\hbar\to 0$, where $\|\cdot\|_{4 \times 4}$ denotes an 
arbitrary matrix norm. We call the quantisations of such symbols
semiclassical observables. These form a subalgebra $\Osc$ in the algebra
of bounded operators on $\mathscr{H}$. As usual, only the self-adjoint 
elements of $\Osc$ are true quantum mechanical observables; they can be 
characterised by symbols taking values in the hermitian $4\times 4$ matrices. 

The crucial point now is to identify a subalgebra $\Oinv$ in $\Osc$ that is 
invariant with respect to the quantum time evolution. We recall that in
the case of scalar operators, such as Schr\"odinger-Hamiltonians on
$L^2(\rz^3)$, the respective algebra $\Osc$ itself is invariant, see e.g.
\cite{Rob87}. The Zitterbewegung of a free relativistic particle, 
however, shows that in the present context this can no longer hold. This 
example also suggests that $\Oinv$ might consist of those semiclassical 
observables that are block-diagonal with respect to the semiclassical 
projections $\hat\Pi^\pm$. In the following we are going to demonstrate 
that this is indeed the case.

Since the necessary constructions take place in terms of symbols, we first 
recall how the algebraic properties of semiclassical observables are
reflected on this level, see e.g. \cite{Fol89,Rob87,DimSjo99}: The product
of two semiclassical observables $\hat B,\hat C\in\Osc$ is a Weyl
operator with a symbol denoted $B\# C(\vx,\vp;\hbar)$. This symbol has an
asymptotic expansion of the type (\ref{eq:ClassSymb})-(\ref{eq:AsympExpand}) 
that reads
\begin{equation*}
 B\# C(\vx,\vp;\hbar) \sim \sum_{j,k,l=0}^\infty\frac{\hbar^{j+k+l}}{j!} 
 \left. \Bigl( \frac{\ui}{2} \bigl( \nabla_x\cdot\nabla_\xi -\nabla_p\cdot
 \nabla_y \bigr) \Bigr)^j B_k(\vp,\vx)\,C_l(\vxi,\vy)
 \right|_{\substack{\vy=\vx\\ \vxi =\vp}} \ .
\end{equation*}
We remark that if the operator $\hat C$ is replaced by the Hamiltonian $\hat H$
a corresponding product formula for the symbol $B\# H(\vx,\vp;\hbar)$ exists, 
with the exception that on the right-hand side of the estimates 
(\ref{eq:AsympExpand}) the order function (\ref{eq:OrderFct}) appears as an 
additional factor. 

We are now in a position to write down the Heisenberg equation of motion 
for the time evolution $\hat B(t)$ of a semiclassical observable in terms of 
symbols,
\begin{equation} 
\label{eq:HeisenbergSymb} 
 \frac{\partial}{\partial t} B(t)(\vx,\vp;\hbar) = 
 \frac{\ui}{\hbar} \bigl( H\# B(t) - B(t)\# H \bigr) (\vx,\vp;\hbar) \ .
\end{equation}
The strategy we follow now is to assume an asymptotic expansion of the
form (\ref{eq:AsympExpand}) for the symbol $B(t)(\vx,\vp;\hbar)$, insert
this into (\ref{eq:HeisenbergSymb}) and solve, if possible, the hierarchy
of equations for the coefficients $B(t)_k(\vx,\vp)$ of $\hbar^k$. A first
obstacle to this procedure arises upon investigating the leading terms
on the right-hand side of (\ref{eq:HeisenbergSymb}),
\begin{equation} 
\label{eq:HeisenbergSymblead} 
 \frac{\partial}{\partial t} B(t) = \frac{\ui}{\hbar}[H, B(t)_0] -
 \frac{1}{2}\bigl( \{H,B(t)_0 \} - \{ B(t)_0,H \} \bigr) +  
 \ui [H,B(t)_1] + O(\hbar) \ .
\end{equation}
Here $[\cdot,\cdot]$ is a matrix commutator and 
\begin{equation*} 
 \{B,C\}(\vx,\vp) = \bigl( \nabla_p B \cdot \nabla_x C - 
 \nabla_x B \cdot \nabla_b C \bigr) (\vx,\vp)
\end{equation*}
denotes the Poisson bracket for matrix valued functions on phase space.
Due to the factor of $1/\hbar$ in the first term on the right-hand side of
(\ref{eq:HeisenbergSymblead}) the proposed strategy can only be consistent if
the matrix valued function $B(t)_0(\vx,\vp)$ is block-diagonal with respect
to the eigenprojections $\Pi_0^\pm(\vx,\vp)$ of the symbol matrix 
$H(\vx,\vp)$. This condition is the lowest semiclassical order of the 
expected restriction on the operators in $\Oinv$.

Necessary and sufficient conditions to be imposed on semiclassical
observables to be in $\Oinv$ can be obtained in any semiclassical order,
if the solvability of the equations (\ref{eq:HeisenbergSymb}) for the 
diagonal and for the off-diagonal blocks of $B(t)$ with respect to the 
projection symbols $\Pi^\pm$ are investigated. This leads to:
\begin{theorem}
\label{Thm:InvObserv}
A semiclassical observable $\hat B\in\Osc$ lies in the invariant algebra
$\Oinv$, if and only if its off-diagonal blocks with respect to the 
semiclassical projections $\hat P^\pm$ are smaller than any power of
$\hbar$,
\begin{equation}
\label{eq:SCLblock}
 \hat P^\pm\hat B\hat P^\mp  = O(\hbar^\infty) \ . 
\end{equation}
\end{theorem}
This statement is a version appropriate for the Dirac equation of a result
contained in \cite{BolGla02}. There one can also find a proof.

The Hamiltonian (\ref{eq:DiracHam}) is not in $\Oinv$, but this is only 
due to our restriction to bounded operators; otherwise, the relations 
(\ref{eq:SemiclCommute}) and (\ref{eq:SemiclResUn}) imply that $\hat H$
indeed possesses the property (\ref{eq:SCLblock}). Furthermore, up to
errors of the order $\hbar^\infty$ the time evolution of the diagonal
blocks $\hat P^\pm\hat B\hat P^\pm$ is governed by the projected 
Hamiltonians $\hat H\hat P^\pm$. An interpretation of 
Theorem~\ref{Thm:InvObserv} now is rather obvious: An observable remains 
semiclassical under the quantum dynamics if it does not contain off-diagonal 
blocks representing an interference of particle and anti-particle dynamics, 
since this has no classical equivalent. Terms exceptional to this must be 
smaller than any power of $\hbar$, indicating that they arise from pure 
quantum effects.

\section{Classical limit of quantum dynamics}
\label{sec:TimeEvol}
 
Given an observable $\hat B\in\Oinv$, we are interested in the connection 
between its time evolution $\hat B(t)$ and appropriate classical dynamics.
The diagonal blocks of $\hat B$ are (approximately) propagated with 
the projected Hamiltonians $\hat H\hat P^\pm$, whose symbols read
\begin{equation*}
 H(\vx,\vp)\,\Pi_0^\pm(\vx,\vp) = h_\pm  (\vx,\vp)\,\Pi_0^\pm(\vx,\vp)
\end{equation*}
to leading order. We therefore expect the equations of motion generated by 
the eigenvalue functions (\ref{eq:DiracSymbEv}),
\begin{equation}
\label{eq:HamiltonEOM}
 \dot{\vx}_\pm(t) = \{ h_\pm, \vx_\pm(t) \} \ , \qquad 
 \dot{\vp}_\pm(t) = \{ h_\pm, \vp_\pm(t) \} \ ,
\end{equation}
to play an important role for the classical limit of the quantum time 
evolution. The solutions to (\ref{eq:HamiltonEOM}) define the Hamiltonian 
flows $\Phi^t_\pm(\vp,\vx)=(\vp_\pm(t),\vx_\pm(t))$, with 
$(\vp_\pm(0),\vx_\pm(0))=(\vp,\vx)$, on phase space. These flows represent
the classical dynamics of relativistic particles (and anti-particles).
Spin is absent from these expressions, reflecting the fact that a 
priori spin is a quantum mechanical concept and its classical counterpart
has to be recovered in systematic semiclassical approximations of the 
quantum system. 

In the present situation the kinematics and the dynamics of 
spin are encoded in the matrix part of the symbols of observables. For an 
observable $\hat B\in\Oinv$ the leading semiclassical order of its symbol 
is composed of the functions
\begin{equation}
\label{eq:PrincipalBlocks}
 \bigl( \Pi_0^\pm\,B_0\,\Pi_0^\pm \bigr) (\vx,\vp)
\end{equation}
taking values in the (hermitian) $4\times 4$ matrices. These matrices act
on the two dimensional subspaces $\Pi_0^\pm(\vx,\vp)\kz^4$ of $\kz^4$, which
can be viewed as Hilbert spaces of a spin 1/2 attached to a classical 
particle or anti-particle, respectively, at the point $(\vx,\vp)$ in phase 
space. In order to work in these two dimensional spaces explicitly, one 
can introduce orthonormal bases $\{e^\pm_1(\vx,\vp),e^\pm_2(\vx,\vp)\}$
for each of them. For convenience the vectors $e^\pm_k(\vx,\vp)\in\kz^4$ 
are chosen as eigenvectors of the symbol matrix $H(\vx,\vp)$ with 
eigenvalues $h_\pm(\vx,\vp)$. Upon expanding vectors from
$\Pi_0^\pm(\vx,\vp)\kz^4$ in these bases one introduces isometries
$V_\pm (\vx,\vp):\kz^2\to\Pi_0^\pm(\vx,\vp)\kz^4$, such that 
$V_\pm (\vx,\vp)V_\pm^\ast (\vx,\vp)=\Pi_0^\pm(\vx,\vp)$ and 
$V_\pm^\ast (\vx,\vp)V_\pm (\vx,\vp)=\eins_2$. A possible choice for 
these isometries is  
\begin{equation*}
\begin{split}
 V_+ (\vx,\vp) &= 
 \frac{1}{\sqrt{2\varepsilon(\vx,\vp)(\varepsilon(\vx,\vp)+mc^2)}}
 \begin{pmatrix} \bigl( \varepsilon(\vx,\vp)+mc^2 \bigr)\eins_2 \\ 
 \bigl( c\vp-e\vA(\vx) \bigr)\cdot\vsigma \end{pmatrix} \ ,  \\
V_- (\vx,\vp) &= 
 \frac{1}{\sqrt{2\varepsilon(\vx,\vp)(\varepsilon(\vx,\vp)+mc^2)}}
 \begin{pmatrix} \bigl( c\vp-e\vA(\vx) \bigr)\cdot\vsigma \\ 
 - \bigl( \varepsilon(\vx,\vp)+mc^2 \bigr)\eins_2 \end{pmatrix} \ .
\end{split}
\end{equation*}
That way the diagonal blocks (\ref{eq:PrincipalBlocks}) can be represented 
in terms of the $2\times 2$ matrices
\begin{equation}
\label{eq:PrincipalBlocks2}
 \bigl( V_\pm^\ast\,B_0\,V_\pm \bigr) (\vx,\vp) \ ,
\end{equation}
which are hermitian if $B_0(\vx,\vp)$ is hermitian on $\kz^4$. 

The matrix valued functions (\ref{eq:PrincipalBlocks2}) can be mapped in a 
one-to-one manner to real valued functions on the sphere $\Stwo$ via
\begin{equation}
\label{eq:PrinBlocksphere}
 b_0^\pm(\vx,\vp,\vn) := \frac{1}{2}\mtr\Bigl( \Delta_{1/2}(\vn)\,
 \bigl( V_\pm^\ast\,B_0\,V_\pm \bigr) (\vx,\vp) \Bigr)\ ,
\end{equation}
where $\vn\in\rz^3$ with $|\vn|=1$ is viewed as a point on $\Stwo$ and
\begin{equation*}
 \Delta_{1/2}(\vn) := 
 \frac{1}{2}\bigl( \eins_2 + \sqrt{3}\;\vn\cdot\vsigma \bigr) 
\end{equation*}
takes values in the hermitian $2\times 2$ matrices, see \cite{VarGra89}. The 
quantisers $\Delta_{1/2}(\vn)$ provide a quantum-classical correspondence 
on the sphere that is covariant with respect to $\SU(2)$-rotations in the 
following sense,
\begin{equation}
\label{eq:CovQuantiser}
 g\,\Delta_{1/2}(\vn)\,g^{-1} =  \Delta_{1/2}\bigl(R(g)\vn\bigr) \ .
\end{equation}
Here $R(g)\in\SO(3)$ is a rotation associated with $g\in\SU(2)$ through
$g\vn\cdot\vsigma g^{-1}=(R(g)\vn)\cdot\vsigma$, see \cite{BolGlaKep01} 
for further details. The semiclassically leading term of a (block-diagonal)
semiclassical observable $\hat B$ can therefore be represented by two
real valued functions on the space $\rz^3\times\rz^3\times\Stwo$. 
Through the relation
\begin{equation*}
 \bigl( V_\pm^\ast\,B_0\,V_\pm \bigr) (\vx,\vp) = \int_{\Stwo} 
 b_0^\pm(\vx,\vp,\vn) \, \Delta_{1/2}(\vn) \ \ud\vn \ ,
\end{equation*}
where $\ud\vn$ is the normalised area form on $\Stwo$, the matrix valued 
expressions (\ref{eq:PrincipalBlocks2}) are unambiguously recovered from the 
functions (\ref{eq:PrinBlocksphere}) \cite{BolGlaKep01}. 

The point $\vn\in\Stwo$ that arises as an additional variable in a general
classical observable (\ref{eq:PrinBlocksphere}) can be viewed as a classical 
equivalent of spin. This interpretation is suggested by the fact that the
classical observable (\ref{eq:PrinBlocksphere}) associated with quantum spin, 
\begin{equation*}
 \hat B=\frac{\hbar}{2} \, \Sigma_k  \qquad\text{with}\qquad \Sigma_k :=
 \begin{pmatrix} \sigma_k & 0 \\ 0 &  \sigma_k  \end{pmatrix} \ ,
\end{equation*}
can be calculated as 
\begin{equation*}
\begin{split}
 b_0^\pm(\vx,\vp,\vn) 
  &=  \frac{\hbar}{2}\,\mtr\Bigl(\Delta_{1/2}(\vn)\,
      V_\pm^\ast(\vx,\vp)\,\Sigma_k\,V_\pm(\vx,\vp) \Bigr) \\
  &= \sqrt{\tfrac{1}{2}\bigl(\tfrac{1}{2}+1\bigl)}\,\hbar\, n_k \ .
\end{split}
\end{equation*}
Up to its normalisation, which depends on $\hbar$ and the spin quantum number
$s=1/2$, the point $\vn\in\Stwo$ can thus be considered as a classical spin. 
The latter is therefore represented on its natural phase space $\Stwo$, and 
the space $\rz^3\times\rz^3\times\Stwo$ on which the classical observables 
(\ref{eq:PrinBlocksphere}) are defined can be viewed as the combined phase 
space of the translational and the spin degrees of freedoms.

We now want to identify the classical dynamics of both the translational
and the spin degrees of freedom that emerge in the semiclassical limit
of the quantum dynamics generated by the Hamiltonian $\hat H$. To this end
we recall that $\Oinv$ was defined to contain those semiclassical observables
$\hat B\in\Osc$ whose time evolutions 
$\hat B(t) =\hat U(t)^\ast\hat B\hat U(t)$ are again semiclassical 
observables. Hence $\hat B(t)$ is a Weyl operator of the form 
(\ref{eq:WeylQuant}) with symbol $B(t)(\vx,\vp;\hbar)$ that possesses an 
asymptotic expansion of the type (\ref{eq:ClassSymb}). The leading term
$B(t)_0(\vx,\vp)$ in this expansion then yields the classical time evolution
of both types of degrees of freedom.  
\begin{theorem} 
\label{thm:Egorov}
When $\hat B\in\Oinv$ the semiclassically leading term in the asymptotic 
expansion of the symbol of its time evolution $\hat B(t)$ reads
\begin{equation} 
\label{eq:EgorovPSymb} 
 B(t)_0(\vx,\vp) = \sum_{\nu\in\{+,-\}} d_\nu^\ast(\vx,\vp,t) \, 
 (\Pi_0^\nu B_0 \Pi_0^\nu)\bigl(\Phi^t_\nu (\vx,\vp)\bigr) \, 
 d_\nu (\vx,\vp,t) \ .
\end{equation}
The unitary $4\times 4$ matrices $d_\pm(\vx,\vp,t)$ are determined by the
transport equations
\begin{equation} 
\label{eq:SpinTransEq}
 \dot d_\pm(\vx,\vp,t) + \ui H_\pm\bigl(\Phi^t_\pm(\vx,\vp)\bigr) \, 
 d_\pm(\vp,\vx,t) = 0 \ , \qquad d_\pm(\vp,\vx,0) = \eins_4\ ,
\end{equation}
where the effective spin-Hamiltonians $H_\pm$ are defined as
\begin{equation} 
\label{eq:EffHamilton} 
 H_\pm := -\ui [\Pi_0^\pm,\{h_\pm,\Pi_0^\pm\}] + 
 \frac{\ui}{2}\bigl( h_\pm\Pi_0^\pm\{\Pi_0^\pm,\Pi_0^\pm\}\Pi_0^\pm +
 \Pi_0^\pm\{\Pi_0^\pm,H-h_\pm\Pi_0^\pm\}\Pi_0^\pm \bigr) \ .
\end{equation}
\end{theorem}
We refrain from giving explicit expressions for the $4\times 4$ effective
spin-Hamiltonians (\ref{eq:EffHamilton}) here since below we will only work 
in a $2\times 2$ representation; instead we refer to \cite{Spo00}.
A statement of the type made in Theorem \ref{thm:Egorov} is usually called 
an Egorov theorem \cite{Ego69}. The present version is covered by the 
Egorov theorem in \cite{BolGla02}, where one can also find a proof.

The $4\times 4$ matrices $d_\pm(\vx,\vp,t)$ can be shown to map the two 
dimensional subspaces $\Pi_0^\pm(\vx,\vp)\kz^4$ of $\kz^4$ unitarily to the 
propagated subspaces $\Pi_0^\pm(\Phi^t_\pm(\vx,\vp))\kz^4$ \cite{BolGla02}. 
They hence describe the transport of the (anti-) particle spin along the 
classical trajectories $(\vx_\pm(t),\vp_\pm(t))$. If one prefers to work
in the orthonormal eigenbases $\{e^\pm_1(\vx,\vp),e^\pm_2(\vx,\vp)\}$ 
of the spaces $\Pi_0^\pm(\vx,\vp)\kz^4$ one can introduce the unitary
$2\times 2$ matrices
\begin{equation*} 
 D_\pm (\vx,\vp,t) := V_\pm^\ast \bigl(\Phi^t_\pm(\vx,\vp)\bigr) \,
 d_\pm (\vx,\vp,t) \, V_\pm (\vx,\vp) \ ,
\end{equation*}
which are determined by the equations
\begin{equation} 
\label{eq:2x2SpinTransEq}
 \dot D_\pm(\vx,\vp,t) + \frac{\ui}{2}\vC_\pm\bigl(\Phi^t_\pm(\vx,\vp)\bigr)
 \cdot\vsigma \, D_\pm(\vp,\vx,t) = 0 \ , \qquad D_\pm(\vp,\vx,0) = \eins_2\ ,
\end{equation}
following from (\ref{eq:SpinTransEq}). The transformed effective 
spin-Hamiltonians $\vC_\pm\cdot\vsigma/2$ derive from (\ref{eq:EffHamilton}) 
and can be expressed in terms of the electromagnetic fields through
\begin{equation*} 
 \vC_\pm (\vx,\vp) = \mp\frac{ec}{\varepsilon(\vx,\vp)} \left(
 \vB(\vx) \pm \frac{1}{\varepsilon(\vx,\vp)+mc^2}\Bigl( c\vE(\vx) \times 
 \bigl( c\vp - e\vA(\vx) \bigr) \Bigr) \right)  \ ,
\end{equation*}
where $\varepsilon(\vx,\vp)$ is defined in (\ref{eq:OrderFct}). Since 
therefore the effective spin-Hamiltonians are hermitian and traceless 
$2\times 2$~matrices, the solutions $D_\pm$ of (\ref{eq:2x2SpinTransEq}) are 
in $\SU(2)$. Related results have previously been obtained in a WKB-type 
situation \cite{RubKel63,BolKep98,BolKep99a} and in the context of a 
semiclassical propagation of Wigner functions \cite{Spo00}. 

The transport matrices $D_\pm(\vx,\vp,t)$ carry two-component spinors along 
the trajectories $\Phi^t_\pm(\vx,\vp)$ and induce a classical spin dynamics
along (anti-) particle trajectories. These combined classical dynamics can
be recovered upon representing the leading symbol (\ref{eq:EgorovPSymb})
in terms of the functions (\ref{eq:PrinBlocksphere}) on the combined
phase space,
\begin{equation*} 
\begin{split}
 b(t)_0^\pm (\vx,\vp,\vn) 
 &= \mtr\Bigl( \Delta_{1/2}(\vn)\, \bigl( V_\pm^\ast\,B(t)_0\,V_\pm \bigr) 
    (\vx,\vp) \Bigr)  \\
 &= \mtr\Bigl( D_\pm (\vx,\vp,t)\,\Delta_{1/2}(\vn)\,D^\ast_\pm (\vx,\vp,t)
    \,\bigl( V^\ast_\pm\,B_0\,V_\pm \bigr) (\Phi^t_\pm(\vx,\vp)) \Bigr) \\
 &= b_0^\pm \bigl( \Phi^t_\pm(\vx,\vp),R(D_\pm (\vx,\vp,t))\vn \bigr) \ ,
\end{split}
\end{equation*}
where in the last line the covariance (\ref{eq:CovQuantiser}) has been used.
Hence, in leading semiclassical order the dynamics of an observable 
$\hat B\in\Oinv$ can be expressed in terms of the classical time evolutions
\begin{equation} 
\label{eq:ClassDyn} 
 (\vx,\vp) \mapsto \Phi^t_\pm(\vx,\vp) \qquad\text{and}\qquad
 \vn \mapsto \vn_\pm(t) = R(D_\pm (\vx,\vp,t))\vn \ .
\end{equation}
The spin motion on the sphere thus emerging obeys the equation
\begin{equation} 
\label{eq:ThomasPrec} 
 \dot{\vn}_\pm (t) = \vC_\pm \bigl( \Phi^t_\pm (\vx,\vp) \bigr)
 \times \vn_\pm (t) \ ,\qquad \vn_\pm (0)=\vn \ ,
\end{equation}
that is implied by (\ref{eq:2x2SpinTransEq}). These spin dynamics exactly 
coincide with the Thomas precession that was derived in a purely classical 
context \cite{Tho27}.

Both types of time evolutions in (\ref{eq:ClassDyn}) can be combined to
yield the dynamics
\begin{equation} 
\label{eq:SkewFlow} 
 (\vx,\vp,\vn) \mapsto Y_\pm^t (\vx,\vp,\vn) :=
 \bigl( \Phi^t_\pm(\vx,\vp),R(D_\pm (\vx,\vp,t))\vn \bigr)
\end{equation}
on the phase space $\rz^3\times\rz^3\times\Stwo$. The properties
\begin{equation*} 
 Y_\pm^{t=0} (\vx,\vp,\vn) = (\vx,\vp,\vn) \qquad\text{and}\qquad
 Y_\pm^{t'+t}(\vx,\vp,\vn) = Y_\pm^{t'}\bigl(Y_\pm^t (\vx,\vp,\vn)\bigr)
\end{equation*}
can easily be established, showing that (\ref{eq:SkewFlow}) defines flows
on the combined phase space. These flows are composed of the Hamiltonian
flows $\Phi^t_\pm$ defined in (\ref{eq:HamiltonEOM}) on the ordinary phase 
space $\rz^3\times\rz^3$ and the spin dynamics (\ref{eq:ThomasPrec}) on 
$\Stwo$ that are driven by the Hamiltonian flows. Dynamical systems of this 
type are known as skew-product flows, see e.g. \cite{CorFomSin82}. They leave 
the normalised measures
\begin{equation*} 
 \ud\ell_E^\pm (\vx,\vp) \, \ud\vn = \frac{1}{\vol\Omega_E^\pm}\,
 \delta \bigl( h_\pm (\vx,\vp) - E \bigr)\ \ud x \, \ud p \, \ud\vn
\end{equation*}
on the combined phase space invariant that are products of Liouville measure
$\ud\ell_E^\pm (\vx,\vp)$ on $\Omega_E^\pm$ (microcanonical ensemble) and the 
normalised area measure $\ud\vn$ on $\Stwo$. We hence now conclude that the 
classical limit of the quantum dynamics generated by the Dirac-Hamiltonian 
(\ref{eq:DiracHam}) is given by the two skew-product flows (\ref{eq:SkewFlow})
combining the Hamiltonian relativistic motion of (anti-) particles with the 
spin precession along the (anti-) particle trajectories.

\section{Semiclassical behaviour of eigenspinors}
\label{sec:SCeigen}

In section~\ref{sec:SclCalc} we saw that in general eigenspinors $\psi_n$
of the Dirac-Hamiltonian (\ref{eq:DiracHam}) cannot uniquely be associated
with either particles or anti-particles. For the purpose of semiclassical
studies we therefore prefer to work with the normalised projected
eigenspinors
\begin{equation}
\label{eq:NormProjEigenSpin}
 \phi_n^\pm := \frac{\hat P^\pm\psi_n}{\|\hat P^\pm\psi_n\|} \ .
\end{equation}
Due to the fact that the projectors almost commute with the Hamiltonian 
(\ref{eq:SemiclCommute}), the quantities (\ref{eq:NormProjEigenSpin}) are
almost eigenspinors (quasimodes) of both $\hat H$ and $\hat H\hat P^\pm$,
\begin{equation*}
 \bigl\| \bigl( \hat H-E_n \bigr) \phi^\pm_n \bigr\| = r_n^\pm
 \qquad\text{and}\qquad
 \bigl\| \bigl( \hat H\hat P^\pm-E_n \bigr) \phi^\pm_n \bigr\| = s_n^\pm \ ,
\end{equation*}
with error terms $r_n^\pm$ and $s_n^\pm$ given by
\begin{equation*}
 r_n^\pm = \frac{\| [\hat H,\hat P^\pm]\psi_n \|}{\|\hat P^\pm\psi_n\|}
 \qquad\text{and}\qquad
 s_n^\pm = \frac{\| [\hat H\hat P^\pm,\hat P^\pm]\psi_n \|}
 {\|\hat P^\pm\psi_n\|}\ .
\end{equation*}
Thus, if the norms $\|\hat P^\pm\psi_n\|$ of the projected eigenspinors
are not too small, i.e. if $\|\hat P^\pm\psi_n\|\geq C\hbar^N$ holds for some
$C>0$ and $N<\infty$, the error terms $r_n^\pm$ and $s_n^\pm$ are of size 
$\hbar^\infty$.

This property implies that if the spectra of the operators $\hat H$ and 
$\hat H\hat P^\pm$ are discrete in the intervals $[E_n-r_n^\pm,E_n+r_n^\pm]$ 
and $[E_n-s_n^\pm,E_n+s_n^\pm]$, respectively, these operators each possess 
at least one eigenvalue in the respective interval, see e.g. \cite{Laz93}. 
For $\hat H$ the statement is trivial since $E_n$ is known to be an 
eigenvalue, but regarding $\hat H\hat P^\pm$ it provides new insight. In
particular, the mean spectral densities of the operators $\hat H\hat P^\pm$ 
are in general smaller than that of $\hat H$, see (\ref{eq:Weyl}) below. 
Hence, the eigenvalues $E_n$ of $\hat H$ must cluster in such a way that 
successive intervals $[E_n-r_n^\pm,E_n+r_n^\pm]$ overlap and therefore several
such intervals share eigenvalues of $\hat H\hat P^\pm$. That way the 
separation of eigenvalues of $\hat H$ is linked with the norms of projected 
eigenspinors $\hat P^\pm\psi_n$. Suppose, e.g., all norms were bounded from 
below by $\|\hat P^\pm\psi_n\|\geq C\hbar^N$, such that 
$r_n^\pm,s_n^\pm = O(\hbar^\infty)$, then sufficiently many eigenvalues 
$E_n$ have to cluster on a scale $O(\hbar^\infty)$. On the other hand, if
the separations of neighbouring eigenvalues of $\hat H$ were known to be 
bounded from below by $\hbar^N$, a certain number of eigenspinor-projections
would have to possess norms of size $\hbar^\infty$ in order to produce
sufficiently large errors $r_n^\pm$.

More quantitative statements can be made on the ground of Weyl's law for
the numbers $N_{E,\omega}$ and $N_{E,\omega}^\pm$ of eigenvalues the 
operators $\hat H$ and $\hat H\hat P^\pm$, respectively, possess in the 
interval $[E-\hbar\omega,E+\hbar\omega]$. Namely, as $\hbar\to 0$
\begin{equation}
\label{eq:Weyl}
\begin{split}
 N_{E,\omega}     &\sim\frac{2\omega}{\pi}\,
                    \frac{\vol\Omega_E^+ + \vol\Omega_E^-}{(2\pi\hbar)^2}\ ,\\
 N_{E,\omega}^\pm &\sim\frac{2\omega}{\pi}\,
                     \frac{\vol\Omega_E^\pm}{(2\pi\hbar)^2} \ .                
\end{split}
\end{equation}
Thus, if at energy $E$ both energy shells $\Omega_E^\pm$ in phase space have 
positive volumes, the ratios $N^\pm_{E,\omega}/N_{E,\omega}$ governing the 
previous discussion are semiclassically determined by the relative fraction 
of the volumes of the associated energy shells. Moreover, a version of
Weyl's law that includes expectation values of operators (Szeg\"o limit
formula) yields \cite{BolGla02}
\begin{equation}
\label{eq:projSzegoe}
 N_{E,\omega}^\pm \sim \sum_{E-\hbar\omega\leq E_n\leq E+\hbar\omega}
 \| \hat P^\pm\psi_n \|^2 \ .
\end{equation}
This relation may suggest two (extreme) scenarios: 
\begin{itemize}
\item[(i)]
Roughly $N_{E,\omega}^\pm$ of the projected eigenspinors $\hat P^\pm\psi_n$ 
are close to $\psi_n$ and the remaining projections are semiclassically small. 
\item[(ii)] 
The projections $\hat P^\pm\psi_n$ equidistribute on both energy shells in 
the sense that each $\| \hat P^\pm\psi_n \|^2$ is approximately 
$N_{E,\omega}^\pm/N_{E,\omega}$. 
\end{itemize}
On the ground of the existing knowledge one
cannot exclude either alternative, nor a mixture of both. However, a
comparison with symmetric vs. asymmetric double-well potentials, where an
eigenfunction either localises in one well (asymmetric case, compare (i)),
or has noticeable projections to both wells (symmetric case, compare (ii)),
suggests that in generic cases scenario (i) should be expected.

What is possible, though, is to estimate the number of projected eigenspinors
with norms that are bounded from below,
\begin{equation*}
 N^\pm_\delta := \# \{ n;\ E-\hbar\omega\leq E_n\leq E+\hbar\omega,\ 
 \| \hat P^\pm\psi_n \|^2 \geq\delta \} \ .
\end{equation*}
From (\ref{eq:Weyl}) and (\ref{eq:projSzegoe}) it follows that 
\begin{equation*}
 \lim_{\hbar\to 0}\frac{N^\pm_\delta}{N_{E,\omega}} \geq 
 \frac{\vol\Omega_E^\pm}{\vol\Omega_E^+ + \vol\Omega_E^-} - \delta \ .
\end{equation*}
Therefore, if $\delta$ is sufficiently small (independent of $\hbar$)
the right-hand side is positive and thus a finite portion of the projected 
eigenspinors have semiclassically non-vanishing norms. This observation
allows to take the associated normalised projected eigenspinors 
(\ref{eq:NormProjEigenSpin}) into account for the following considerations. 

We now explore the consequences an ergodic behaviour of the classical 
skew-product dynamics $Y^t_\pm$ exerts on the projected eigenspinors. In
an analogous situation for Schr\"odinger-Hamiltonians the ergodicity of the
associated Hamiltonian flow implies that the Wigner transforms of almost
all eigenfunctions equidistribute on the corresponding energy shell.
This result, known as quantum ergodicity, goes back to Shnirelman \cite{Shn74}
and has been fully proven in \cite{Zel87,Col85,HelMarRob87}. In the case
under study we now suppose that the energy $E$ lies in an interval 
$(E_-,E_+)$, in which the spectrum of the Dirac-Hamiltonian is discrete. 
Moreover, at least one of the classical energy shells $\Omega_E^+$ and 
$\Omega_E^-$ shall be non-empty and the periodic orbits of the Hamiltonian 
flows $\Phi^t_\pm$ shall be of measure zero on $\Omega_E^\pm$. If then 
$Y^t_\pm$ is ergodic on $\Omega_E^\pm\times\Stwo$, the time average of a 
classical observable of the type (\ref{eq:PrinBlocksphere}) equals its 
phase-space average,
\begin{equation*}
 \lim_{T\to\infty}\frac{1}{T}\int_0^T b_0^\pm \bigl( Y^t_\pm (\vx,\vp,\vn)
 \bigr)\ \ud t = \int_{\Omega_E^\pm}\int_{\Stwo} b_0^\pm (\vx',\vp',\vn')
 \ \ud\ell_E^\pm (\vx',\vp') \, \ud\vn' =: M_E^\pm\bigl(b_0^\pm\bigr) \ ,
\end{equation*}
for almost all initial conditions $(\vx,\vp,\vn)$. In this setting the main 
result on quantum ergodicity of projected eigenspinors is
(for a full proof see \cite{BolGla02}):
\begin{theorem}
\label{Thm:QE}
Under the conditions stated above, in particular if the skew-product flow
$Y^t_\pm$ is ergodic on $\Omega_E^\pm\times\Stwo$, in every sequence
$\{ \phi_n^\pm \}_{n\in\nz}$ of normalised projected eigenspinors with
associated eigenvalues $E_n\in [E-\hbar\omega,E+\hbar\omega]$ and 
$\|\hat P^\pm\psi_n \|^2\geq\delta$ there exists a subsequence  
$\{ \phi_{n_\alpha}^\pm \}_{\alpha\in\nz}$ of density one, i.e.
\begin{equation*}
 \lim_{\hbar\to 0}
 \frac{\#\{\alpha;\ \|\hat P^\pm\psi_{n_\alpha} \|^2\geq\delta\}}
 {\#\{n;\ \|\hat P^\pm\psi_n \|^2\geq\delta\}} =1 \ ,
\end{equation*}
such that for every semiclassical observable $\hat B\in\Osc$
\begin{equation}
\label{eq:QE}
 \lim_{\hbar\to 0}\,\langle\phi^\pm_{n_\alpha},\hat B\phi^\pm_{n_\alpha}\rangle
 = M_E^\pm\bigl(b_0^\pm\bigr) \  .
\end{equation}
The density-one subsequence $\{ \phi_{n_\alpha}^\pm \}_{\alpha\in\nz}$ can 
be chosen independent of the observable.
\end{theorem}
The statement of this theorem, that in case of classical ergodicity quantum 
mechanical expectation values converge to a classical ergodic mean, can be
rephrased in terms of more explicit phase-space lifts. To this end one
first introduces the matrix valued Wigner functions
\begin{equation*}
 W[\psi](\vx,\vp) := \int_{\rz^3}\ue^{-\frac{\ui}{\hbar}\vp\cdot\vy}\,
 \overline{\psi}\bigl( \vx-\tfrac{1}{2}\vy \bigr)\otimes
 \psi\bigl( \vx+\tfrac{1}{2}\vy \bigr) \ \ud y 
\end{equation*}
associated with any $\psi\in L^2(\rz^3)\otimes\kz^4$. Following the
prescription (\ref{eq:PrinBlocksphere}) this $4\times 4$ matrix
can be converted into scalar form through
\begin{equation*}
 w_\pm [\psi](\vx,\vp,\vn) := \frac{1}{2}\mtr\Bigl( \Delta_{1/2}(\vn)\,
 \bigl( V_\pm^\ast\,W[\psi]\,V_\pm \bigr) (\vx,\vp) \Bigr) \ .
\end{equation*}
Due to the projections inherent in the expectation value on left-hand side
of (\ref{eq:QE}) only one diagonal block of the observable contributes,
so that this expectation value can be reformulated as
\begin{equation*}
\begin{split}
 \langle\phi^\pm_{n_\alpha},\hat B\phi^\pm_{n_\alpha}\rangle 
 &= \frac{1}{(2\pi\hbar)^3}\iint\mtr\bigl( W[\phi^\pm_{n_\alpha}] \,
    P^\pm \# B \# P^\pm \bigr)(\vx,\vp) \ \ud x\,\ud p \\
 &= \frac{1}{(2\pi\hbar)^3}\iint\mtr\Bigl( \bigl( V_\pm^\ast\,
    W[\phi^\pm_{n_\alpha}]\,V_\pm \bigr) \bigl( V_\pm^\ast\,B_0\,V_\pm
    \bigr)(\vx,\vp) (1+O(\hbar)) \Bigr) \ \ud x\,\ud p \\
 &= \frac{1}{(2\pi\hbar)^3}\iiint \bigl( w_\pm [\phi^\pm_{n_\alpha}]
    (\vx,\vp,\vn)\,b_0^\pm (\vx,\vp,\vn)(1+O(\hbar))\bigr)\ 
    \ud x\,\ud p\,\ud\vn \ .
\end{split}
\end{equation*}
The principal result (\ref{eq:QE}) of quantum ergodicity can hence be read
as saying that
\begin{equation*}
 \lim_{\hbar\to 0}\frac{w_\pm [\phi^\pm_{n_\alpha}](\vx,\vp,\vn)}
 {(2\pi\hbar)^3} = \frac{1}{\vol\Omega_E^\pm}\,\delta\bigl( 
 h_\pm(\vx,\vp) - E \big) \ .
\end{equation*}
This relation has to be understood in a weak sense, i.e. after integration
with a symbol $b_0^\pm (\vx,\vp,\vn)$. Hence, the scalar Wigner transforms
of projected eigenspinors become semiclassically equidistributed on the
associated phase space $\Omega_E^\pm\times\Stwo$ once the classical time 
evolution $Y^t_\pm$ is ergodic on this space.

As the discussion at the beginning of this section shows, the difficulties
with statements about genuine eigenspinors derive from the coexistence of
the particle and anti-particle subspaces; and because interactions between 
these subspaces on a scale $\hbar^\infty$ cannot be controlled within the 
present setting. However, if one of the energy shells $\Omega_E^\pm$ were 
empty there is only one classical manifold onto which phase-space lifts of 
eigenspinors could condense, namely $\Omega_E^\mp\times\Stwo$. In such a 
case, say when $\Omega_E^-$ is empty, the statement of Theorem~\ref{Thm:QE} 
applies to a density-one subsequence $\{\psi_{n_\alpha}\}$ of eigenspinors 
themselves, such that
\begin{equation}
\label{eq:QE2}
 \lim_{\hbar\to 0}\,\langle\psi_{n_\alpha},\hat P^+\hat B\hat P^+
 \psi_{n_\alpha}\rangle = M_E^\pm\bigl(b_0^\pm\bigr) \qquad\text{and}\qquad
 \lim_{\hbar\to 0}\,\langle\psi_{n_\alpha},\hat P^-\hat B\hat P^-
 \psi_{n_\alpha}\rangle = 0 \  .
\end{equation}
A corresponding statement then holds for the associated scalar Wigner 
transforms $w_\pm [\psi_{n_\alpha}]$. It has been mentioned previously that
the restriction of having only one non-empty energy shell at energy $E$ 
requires potentials that do not vary too strongly; e.g., $\phi(\vx)$ must
not vary as much as $2mc^2$. This condition is fulfilled in many physically
relevant situations, in which (\ref{eq:QE2}) therefore applies if only
the classical particle dynamics $Y^t_+$ are ergodic.

\vspace*{0.5cm}
\subsection*{Acknowledgment}
Financial support by the Deutsche Forschungsgemeinschaft (DFG) 
under contracts no. Ste 241/15-1 and /15-2 is gratefully acknowledged.

{\small
\bibliographystyle{amsalpha}
\bibliography{literatur}} 

\end{document}